\documentclass[journal]{IEEEtran}
\usepackage{graphicx}
\usepackage{subfigure}
\usepackage{booktabs}  
\usepackage{multirow}
\usepackage{amsmath}
\usepackage{bbding}
\usepackage{amsmath}
\usepackage{stfloats} 
\usepackage{algorithm} 
\usepackage{algorithmic} 
\usepackage{amssymb}
\usepackage{cite}
\usepackage{CJKutf8} 
\usepackage{cite}
\begin{document}
\begin{CJK}{UTF8}{gkai}  
\title{Design and Implementation of Emergency Simulated Lighting System Based on Tello UAV}
\author{Yexin~Pan,
    Yong~Xu,
    Bo~Ma,
    and Chuanhuang~Li
\thanks{Yexin~Pan, Yong~Xu, Bo~Ma,
and Chuanhuang~Li are with the School of Information and Electronic Engineering, Zhejiang Gongshang University, Hangzhou 310018, China.}
\thanks{Yong Xu is the corresponding author.}}

\maketitle
\begin{abstract}
In recent years, with the increasing maturity of UAV technology, the application of UAV in the civilian field has seen explosive growth due to their low cost, high flexibility, and wide adaptability. In order to address the drawbacks of current tethered UAV lighting, which necessitates manual operation and coordination with tethered cables, this paper presents a rapid reaction and autonomous deployment emergency lighting system prototype based on Tello UAV. The system design is divided into three modules. First, the lighting module has designed the protogype lighting extensions for the Tello UAV. By selecting and installing LED light sources reasonably, the stability of the UAV's takeoff, landing, and flight is not affected, while ensuring good lighting effects; Second, the addressing module extracts the optimization objective function based on the optimization deployment requirements of UAV, and uses simulated annealing algorithm to iteratively optimize for different distribution scenarios and user needs, calculating the optimal deployment location of UAV and planning flight missions for UAV. Third, the flight control module has designed a specialized command control framework based on Tello UAV's API, which converts the planned flight path into command statements, forms flight text, and controls the flight of unmanned aerial vehicles accordingly. The test results of the emergency lighting system show that the system can respond quickly to both densely distributed and sparsely distributed user scenarios, deploying UAV to the best service locations, and providing high-quality services to users. This work represents an early attempt of Python-based real-world UAV lighting system prototype design to associate the UAV hardware and Python algorithms. 
\end{abstract}

\begin{IEEEkeywords}
UAV, Illumination System, Automatic deployment, Path planning.
\end{IEEEkeywords}

\section{Introduction}
\IEEEPARstart{W}{ith} the progress of industrialization and modernization, people are facing more and more safety and environmental issues. Moreover, the probability of unexpected incidents has greatly increased. The need for emergency lighting is growing linearly in order to guarantee both people's safety and the security of property in emergency situations \cite{zhaoming}. Furthermore, people have increased expectations for the design and caliber of emergency lighting as a result of the complexity of emergency situations and the strengthening of laws, regulations, and safety standards. Higher standards have sped up technological advancement in the emergency lighting sector. Lighting technology has advanced significantly in both the light source and power supply departments \cite{zhaoming2}. However, traditional iterative upgrades are getting more and more challenging to meet needs in the real world because of limited lighting range. Therefore, as a crucial component of emergency lighting, the lighting carrier bearing the lighting source has turned into the area where people seek to make breakthroughs. As a result, people have turned their attention to UAV. 

Due to their adaptability, mobility, and convenience, unmanned aerial vehicles (UAV) are predicted to develop as a new kind of emergency lighting carrier. With the further maturity of technologies such as batteries and chips, as well as the rapid development and popularization of 5G communication technology, the technology and overall manufacturing process of UAV have also entered a period of rapid development \cite{wu2022amassing}. In terms of battery technology, Modern UAV can now use lighter, smaller, higher capacity lithium-ion batteries thanks to advances in lithium battery technology, which have significantly increased UAV' range and flight efficiency. At the same time, new sustainable energy batteries such as solar rechargeable batteries and fuel cells have also emerged \cite{chang2022assessment}, providing more options for long-duration UAV flights. In terms of chips and hardware technology, the flight control chips and sensors used in UAV have been greatly improved, such as integrated positioning and navigation systems (INS) composed of inertial navigation systems (IMU), global satellite navigation systems (GNSS), barometers, gyroscopes, accelerometers, and digital signal processors, which can improve the accuracy and reliability of flight control and makes the flight of UAV smoother, more reliable, and safer. Meanwhile, the improvement and enhancement of manufacturing processes have also made the types of UAV increasingly diverse \cite{chamola2021comprehensive}, such as vertical takeoff and landing UAV, fixed wing UAV, multi rotor UAV, etc., with different types having different characteristics and applicable scenarios. The rapid development of UAV has made them no longer limited to the military industry and began to be popularized in the civilian field. With the characteristics of low usage and maintenance costs, high flexibility, and wide adaptability, the application of UAV in the civilian field has shown a trend of explosive development in recent years, with aerial photography, plant protection, security, logistics, emergency response, and inspection as its main application areas \cite{aa,bb,cc,dd,ee}. UAV can fully utilize their distinct benefits when used as a carrier for emergency lights, and this is a disruptive advancement for the emergency lighting. ln this study, an independent emergency lighting system for UAVs, particularly the Tello UAV, is introduced. The system is composed of three distinct modules: lighting, addressing, and flight control. The system has undergone rigorous testing and has demonstrated successful outcomes, signifying a promising step forward in integrating Python algorithms with UAV hardware.


\subsection{Related Work}
The traditional lifting pole emergency lighting method has been developed to a relatively complete extent. Its overall system consists of searchlights, lifting pole, and power supply device. The structure is relatively simple and can be moved and deployed through the bottom wheel group. Considering safety, the height of the lifting pole is generally not higher than 5 meters, and the power supply module at the bottom may increase the counterweight according to the situation. The lifting pole searchlights come in a variety of sizes—from tiny handheld models to sizable ones mounted on vehicles, powered by lithium batteries or diesel generators. However, the lifting pole is difficult to make significant breakthroughs in the height and range of lighting due to the limitations of its own conditions. Moreover, the fixed set of system devices is always difficult to achieve portability in size, and the effect of mobile deployment is poor. It has fallen behind the times and can no longer meet the rapidly changing needs of reality. With the development of more effective UAV illumination, the lifting pole searchlight's application situations have shrunk to mostly small-scale or interior emergencies.
\par
Currently, the relatively mature application of UAV lighting is tethered UAV. The tethered UAV system mainly consists of the tethered UAV platform, data link, and ground equipment. The tethered UAV platform mainly includes UAV body, expansion equipment, etc. The data link, which serves as the connection between the tethered UAV platform and ground equipment, typically consists of an airborne data terminal and a ground data terminal. It is used for a variety of tasks, such as positioning and tracking the tethered UAV, transmitting task payload data, and remote telemetry of the airborne equipment's working parameters and the UAV's flight parameters. The ground equipment consists of a ground power supply, tethered cable, cable retraction and release device, and ground control unit.

\begin{table*}[ht]
\renewcommand\arraystretch{1.1}
\label{table1}
\caption{Path Planning Algorithm}
\scalebox{1.2}{
\begin{tabular}{|c|c|c|c|}
\hline
\textbf{Classification}                                                                          & \textbf{Algorithm} & \textbf{Advantage}                                                                                         & \textbf{Shortcoming}                                                                                                         \\ \hline
\multirow{3}{*}{\begin{tabular}[c]{@{}c@{}}\\Tradition\\ algorithm\end{tabular}}                & A*\cite{A*}                 & Direct search, fast                                                                                        & \begin{tabular}[c]{@{}c@{}}Path planning is not smooth,\\    low efficiency in complex path\end{tabular}                     \\ \cline{2-4} 
                                                                                                 & APF\cite{APF}                & \begin{tabular}[c]{@{}c@{}}Good real-time performance,\\    smooth path\end{tabular}                       & \begin{tabular}[c]{@{}c@{}}Easy to fall into local optimality, \\ endpoint may not be reachable\end{tabular}                 \\ \cline{2-4} 
                                                                                                 & RRT\cite{RRT}                & \begin{tabular}[c]{@{}c@{}}No need for environment modeling,\\  fast computing speed\end{tabular}          & \begin{tabular}[c]{@{}c@{}}Randomness, difficult to obtain   \\ high quality solutions\end{tabular}                          \\ \hline
\multirow{4}{*}{\begin{tabular}[c]{@{}c@{}}\\\\Intelligent \\ optimization\\ algorithm\end{tabular}} & GA\cite{GA}                 & \begin{tabular}[c]{@{}c@{}}Fast heuristic search with good\\ universality\end{tabular}                     & \begin{tabular}[c]{@{}c@{}}Easy to mature early and fall\\    into local optima\end{tabular}                                 \\ \cline{2-4} 
                                                                                                 & ACO\cite{ACO}                & \begin{tabular}[c]{@{}c@{}}Strong global search ability,\\ robustness\end{tabular}                         & \begin{tabular}[c]{@{}c@{}}Sensitive to initial information\\    and prone to falling into local optima\end{tabular}         \\ \cline{2-4} 
                                                                                                 & CS\cite{CS}                 & \begin{tabular}[c]{@{}c@{}}Few parameters, easy to implement,\\   strong optimization ability\end{tabular} & \begin{tabular}[c]{@{}c@{}}Slow convergence and poor\\    population diversity in the later stages of evolution\end{tabular} \\ \cline{2-4} 
                                                                                                 & SA\cite{SA}                 & \begin{tabular}[c]{@{}c@{}}Simple calculation, strong universality,\\  robustness\end{tabular}             & \begin{tabular}[c]{@{}c@{}}Slow convergence, solution\\    quality is related to cooling rate\end{tabular}                   \\ \hline
\end{tabular}}
\end{table*}

The development and use of tethered UAV in the field of lighting has reached a certain level. Zhejiang Jikeqiao Intelligent Equipment Co., Ltd. created the GB1801 tethered UAV, which has a 50000 lumen illumination capacity and can run 600 watt LED lights for four hours on a single power supply. Under the coordination of several power sources or the usage of generators, it can give unbroken continuous lighting 24 hours a day. Lights that can illuminate a 3000 square meter area of the site and conduct rescue operations weigh no more than 1.5 kg, and take-up components and single battery components weigh no more than 20 kg. The nighttime building of Xiong'an New Area and the earthquake night rescue in Turkey both benefited greatly from the use of this tethered UAV. The M300 tethered UAV produced by Guilin tethered aviation technology Co., Ltd. can install multiple sets of searchlights, with a maximum luminous flux of 100000 lumens and a hovering time of more than 12 hours. It has been adopted by fire rescue teams and emergency management bureaus in many Chinese provinces and cities, and has provided night lighting support in the search and rescue of China Eastern Airlines' crashed aircraft. In 2022, Maywell Technology Nanjing Co., Ltd. released a tethered UAV based on the DJI Inspire2, which continuously provides power to the UAV through a ground power of 3.5KW and can achieve 24-hour hovering in the air. It has a 500W lighting power, 60000+ lumens, 40° focusing angle, 5000 square meters+lighting area. Moreover, it has a high safety factor, and even if the power is cut off, the seamless switching of the body battery will not cause the UAV to crash.

In general, the overall design of UAV systems represented by tethered UAV is relatively complete, and their application scenarios are also quite diverse. In the future, they will develop towards lighter weight, intelligence, and higher efficiency. However, tethered UAV are limited by tethered cables and manual operations, and many hazardous scenarios may not be able to be approached and deployed at all. Therefore, this paper aims to design a fully automated emergency lighting system, which not only saves human resource costs, but also can handle many scenarios where existing UAV systems cannot serve. The quality of automatic deployment of UAV lighting depends heavily on UAV path planning. The path planning of UAV is affected by various factors and conditions, and different path planning algorithms have advantages and disadvantages, as shown in TABLE I.

AlRaslan Mohammad et al. \cite{alraslan2022UAV} introduced a simple evaluation method, which considered the intersections between sections and obstacles to find a collision free and near-optimal path, and used the parallel genetic algorithm (PGA) model to shorten the execution time. Ali Zain Anwar et al. \cite{ali2021cooperative} combined the maximal-minimum ant colony optimization (MMACO) and Cauchy mutation (CM) operators, and selected CM operators to enhance the MMACO algorithm by comparing and examining the variation trend of the fitness function of the local optimal position and the global optimal position, eliminating the classical ant colony algorithm (ACO) and MMACO Limitations: Slow convergence speed, easy to fall into local optimal problems, and enable the UAV to fly safely in a dynamic environment; Sahu Bandita et al. \cite{sahu2023modified} designed a hybrid algorithm through the efficient implementation of improved cuckoo search, sine and cosine algorithm and particle swarm optimization. By integrating the spawning behavior of cuckoo species, they realized the efficient global search strategy with parameter modification, the local search strategy with particle swarm optimization and the greedy way of sine and cosine algorithm to achieve benefits

\subsection{Motivation and Contribution}
Traditional emergency lighting is limited by hardware constraints, making UAVs the ideal carriers due to their flexibility. Tethered UAVs have emerged as efficient and safe alternatives, but are limited in range and may not be suitable for complex environments or disasters. Despite these limitations, UAVs remain a favorable option for emergency lighting due to their unique advantages. Therefore, the current tethered UAV are more suitable for scenarios with simple tasks and longer lighting durations, such as the night construction of Huoshenshan Hospital and Leishenshan Hospital during the outbreak of the COVID-19 epidemic, which relied on tethered UAV for night lighting. However, in scenarios similar to disasters, timely lighting assistance is particularly important. Grasping the golden time after an accident can often better help users escape danger. Therefore, designing an emergency lighting system that uses UAV clusters combined with LED light sources for automatic deployment can make up for the lack of rapid lighting services in the existing lighting field. In order to achieve the effect of rapid response and automatic deployment, and to avoid additional losses caused by meaningless flights, it is necessary to design an efficient and reasonable emergency lighting system to schedule the UAV cluster, ensuring that the UAV can have reasonable path planning and and provide services in case of user requests. The major contributions of our work are following:
\begin{itemize}
    \item The emergency lighting system for UAVs discussed in this article has the capability to promptly respond to user requests, extract relevant parameters for analysis and computation, automatically control UAV flights, and deliver services without the need for manual intervention or ground equipment constraints. Additionally, for secondary development, the system employs the open-source Tello TT (hereafter referred to as Tello) to work in conjunction with the emergency lighting system.
    \item A lighting module was specifically designed for the Tello UAV, taking into account its unique characteristics. In addition, a corresponding UAV command framework was developed to align with the system objectives and requirements. The optimal deployment location of UAVs was then iteratively optimized using the simulated annealing algorithm. Subsequently, the planned flight path and service tasks were converted into UAV commands within the command framework, resulting in the generation of flight texts. Finally, the system established a connection with the UAV based on the instructions provided in the flight text, enabling precise control over the UAV's flight operations.
\end{itemize}

The rest of this paper is organized as follows. Section \ref{System Analysis and Design} analyzes the system's objectives and designs system functions, Section \ref{System testing and results} demonstrates the effectiveness of each module and the overall system, and Section \ref{CONCLUSION} summarizes this paper and draws conclusions.

\section{System Analysis and Design}\label{System Analysis and Design}
\subsection{System Objective Analysis}
The designed emergency simulation lighting system should be able to quickly plan a reasonable and efficient UAV flight path, ensure that the UAV can provide lighting services for users at the optimal deployment location, and at the same time, The UAV must simultaneously adjust its own height in accordance with user service requirements. On the premise of ensuring the coverage of all users, the hover height should be lowered appropriately to improve the lighting brightness and provide users with higher quality of service (QoS). After planning the flight path of the UAV, the system needs to calculate and comprehensively analyze the loss caused by the UAV's flight and lighting services. If the current UAV cannot meet the total electricity required for subsequent services, the system will dispatch a replacement UAV to continue providing lighting services, and the current UAV will continue to provide service until the charge is down to the minimum needed for return.

\subsection{Construction of System Mathematical Model}
\subsubsection{User Model}
The user model includes location coordinates $(x,y)$ and lighting brightness requirements $\beta$ and the three major parameters of lighting time demand $t_{\rm{user}}$. The user's z-coordinate value is 0, so there is no need for additional definition. The brightness requirements of users are defined as $\beta  \in \left[ {1,2,3} \right]$ , because the brightness demand of users in the real scene generally presents normal distribution, the brightness demand of users is divided into three levels, where 1 is the lighting demand of the highest brightness, 2 is the lighting demand of normal brightness, and 3 is the lighting demand of low brightness. Different brightness demand corresponds to different flight altitude ranges of UAV. The lighting time requirement of users is defined as the time required for UAV to provide lighting services, which is $t_{\rm{user}}$.

\subsubsection{UAV Service Model}
There are five steps in a full UAV service cycle: take-off, flight deployment, hover lighting, return, and landing. The service cycle of the UAV is defined as T, and the expression for T is shown in (1):
\begin{equation}
T = {T_{{\rm{off}}}} + {T_{{\rm{deploy}}}} + {T_{{\rm{light}}}} + {T_{{\rm{back}}}} + {T_{{\rm{land}}}}
\end{equation}

\subsubsection{UAV Lighting Model}
The height of UAV lighting is defined as $h$, the effective lighting angle of the light source is defined as $\alpha$, and the area generated by lighting is defined as $S$. The lighting model is shown in Fig. 1, and the calculation formula for $S$ is shown in (2):
\begin{equation}
S = \pi  \cdot {\left( {h \cdot \tan \alpha } \right)^2}
\end{equation}

\begin{figure}[tbp]
\centering 
\includegraphics[scale=0.4]{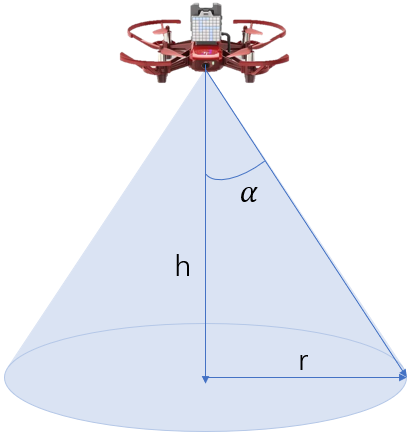}
\caption{UAV Lighting Model}
\label{1}
\end{figure}

\subsubsection{UAV Flight Model}
The UAV's location coordinates are $(X, Y, Z)$, while the user's horizontal distance is D from the UAV. The expression for D is shown in (3):
\begin{equation}
    D = \sqrt {{{\left( {X - x} \right)}^2} + {{\left( {Y - y} \right)}^2}}
\end{equation}

When in a period of $T_{\rm{light}}$, the UAV maintains hovering to provide lighting services. During this period, the displacement of the UAV is 0, and the coordinate positions are $(X, Y, Z)$ unchanged; When in cycles $T_{\rm {off}}$ and $T_{\rm{land}}$, the UAV only moves in the Z-axis direction, with a flight displacement of $\Delta Z$. The coordinates after the UAV displacement are $(X, Y, Z+\Delta Z)$; When the UAV is in cycles $T_{\rm {deploy}}$ and $T_{\rm{back}}$, it moves in the X-axis and Y-axis directions, with flight displacements of $\Delta X$ and $\Delta Y$. The coordinates of the UAV after displacement are $(X+\Delta X, Y+\Delta Y, Z)$, and the expression for the total change of displacement $\Delta sum$ is shown in (4): 
\begin{equation}
\Delta sum = \sqrt {\Delta {X^2} + \Delta {Y^2}}
\end{equation}

The mathematical expression for the change in flight coordinates of a UAV during a complete service cycle is shown in (5):
\begin{small}
\begin{equation}
\left( {X,Y,Z} \right) =\left\{ \begin{array}{l}
\left( {X,Y,Z + \Delta Z} \right), \,\,\,\,\,\,\,\,\,\,\,\,\,\,\,\,\,\,\,\,\,\,\,\,\,\, t \le {T_{{\rm{off}}}}\\
\left( {X + \Delta X,Y + \Delta Y,Z} \right),\,\,\,\,\,\,\,{T_{{\rm{off}}}} < t \le {T_{{\rm{deploy}}}}\\
\left( {X,Y,Z} \right),\,\,\,\,\,\,\,\,\,\,\,\,\,\,\,\,\,\,\,\,\,\,\,\,\,\,\,\,\,\,\,\,\,\,\,\,\,\,\,\,\,\,{T_{{\rm{deploy}}}} < t \le {T_{{\rm{light}}}}\\
\left( {X + \Delta X,Y + \Delta Y,Z} \right),\,\,\,\,\,\,\,{T_{{\rm{light}}}} < t \le {T_{{\rm{back}}}}\\
\left( {X,Y,Z + \Delta Z} \right),\,\,\,\,\,\,\,\,\,\,\,\,\,\,\,\,\,\,\,\,\,\,\,\,\,{T_{{\rm{back}}}} < t \le {T_{{\rm{land}}}}
\end{array} \right.    
\end{equation}
\end{small}

\subsubsection{UAV Power Loss Model}
The power loss of the UAV is defined as E, and the resulting loss comes from the UAV's flight and lighting. $E_{\rm{fly}}$ and $E_{\rm {light}}$ are the definitions for the UAV's flight loss and lighting loss, respectively. The expression for the UAV's power loss E is shown in (6):
\begin{equation}
    E = {E_{{\rm{fly}}}} + {E_{{\rm{light}}}}
\end{equation}

The lighting loss coefficient of the UAV is defined as $\mu $, with a value of 0.2 for $\mu $, and the expression for $E_{\rm{light}}$ is shown in (7):
\begin{equation}
    {E_{{\rm{light}}}} = \mu  \cdot  {T_{{\rm{light}}}}
\end{equation}

The definition of the $E_{\rm{fly}}$ expression for flight loss is based on reference \cite{zeng2019energy}. Firstly, $E_v (t)$ is defined to represent the flight loss of the UAV during each time period, and its expression is shown in (8): 
\begin{small}
\begin{equation}
\begin{array}{l}
{E_v}(t) = {P_0}\left( {\frac{1}{{v(t)}} + \frac{{3v(t)}}{{U_{{\rm{tip}}}^2}}} \right) + {P_1}{\left( {\sqrt {v{{(t)}^{ - 4}} + \frac{1}{{4v_0^4}}}  - \frac{1}{{2v_0^2}}} \right)^{1/2}}\\
\\
\,\,\,\,\,\,\,\,\,\,\,\,\,\,\,\,+ \frac{1}{2}{d_0}\rho sGv{(t)^2}
\end{array}
\end{equation}    
\end{small}

Wherein, ${P_0} = \frac{\delta }{8}\rho sG{\Omega ^3}{R^3}$ and ${P_1} = \left( {1 + k} \right)\frac{{{W^{3/2}}}}{{\sqrt {2\rho G} }}$ are respectively the constant blade power and induced power in hover, $U_{\rm{tip}}$ is the tip speed of rotor blade, $v_0$ is the average rotor induced velocity in hover, $v(t)$ is the flight speed of UAV, $d_0$ and $s$ are respectively the fuselage resistance ratio and rotor fixity (defined as the ratio of total blade area to rotor area), $\rho $ and $G$ are respectively the density of air and rotor disk area, $\Omega $ is the blade angular velocity, $R$ is the rotor radius, $k$ is the correction coefficient for the induced power increment, and $W$ is the weight of the UAV. Therefore, the flight energy consumption of the UAV from the first time slot to the last time slot is expressed as equation (9):
\begin{equation}
    {E_{{\rm{fly}}}} = \sum\nolimits_0^T {{E_v}(t)}
\end{equation}


\subsubsection{Overall System Model}
After receiving a user request, the system will gather the position coordinates, brightness requirements, and time requirements of each user, input the necessary data, and employ the simulated annealing algorithm to iteratively locate the UAV's best deployment location. Then convert the obtained position coordinates into commands and parameters that the UAV can recognize and create a flight text to record. After the flight text is generated, the system will control the UAV's flight accordingly. In accordance with the planned path, the UAV will be deployed to the appropriate location to provide illumination services for users. The UAV must make sure the return power is present while serving users. If the initial departure UAV's power cannot provide sufficient lighting service time for users, the system will dispatch a replacement UAV and the current UAV will return. The overall operation process of the system is shown in Fig. 2.

\begin{figure}[htbp]
\includegraphics[scale=0.343]{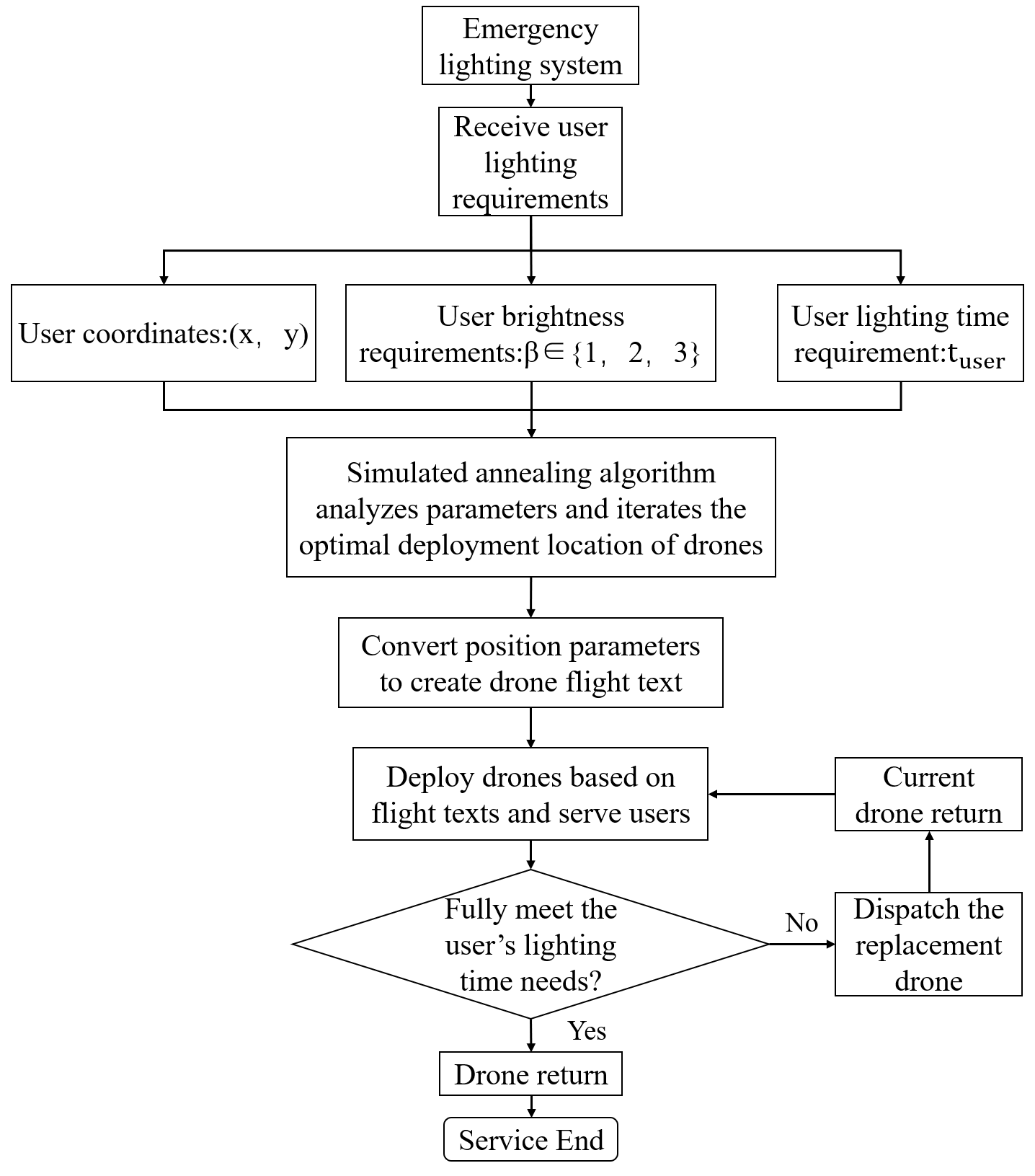}
\caption{Overall Process of Emergency Simulated Lighting System}
\label{2}
\end{figure}

\subsection{Lighting Module Design}
The Tello UAV does not come with a built-in lighting source, hence an external light source must be installed in order to provide lighting. The amount of extra weight the Tello UAV can carry for takeoff is severely constrained due to its diminutive size and 87g weight. Therefore, a reasonable balance must be made between volume, weight, and brightness in the selection of external light sources. The final external light source chosen is a circular LED light source with a 32mm diameter, 15mm thickness, and only 10g weight. It is powered by UAV power, weighs within the Tello UAV's maximum takeoff weight range, and has an effective lighting angle of roughly 30 degrees.

The installation location and quantity of light sources need to be reasonably considered in conjunction with the inherent conditions of the Tello UAV. The bottom view of the Tello UAV is shown in Fig. 3.


\begin{figure}[tbp]
\centering 
\includegraphics[scale=0.3]{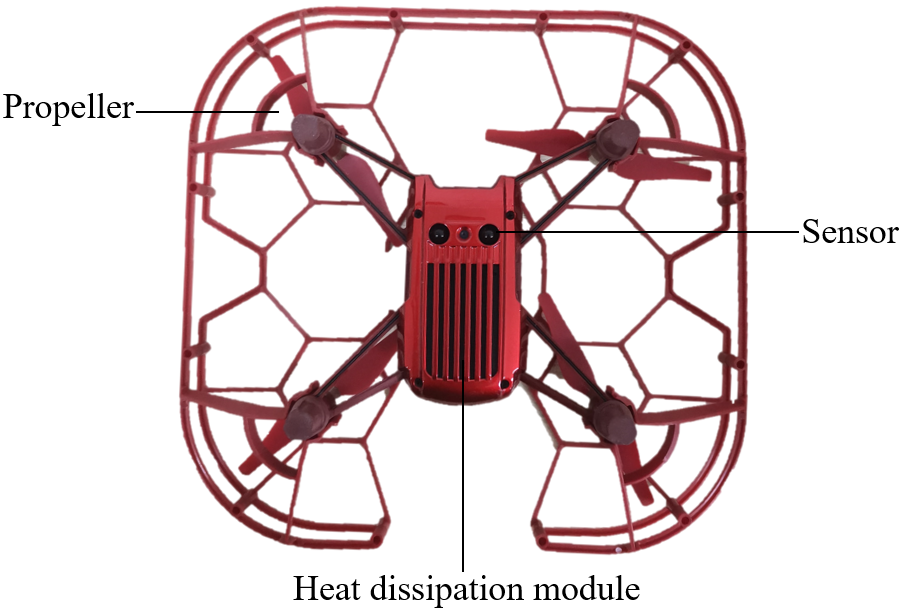}
\caption{Bottom View of Tello}
\label{3}
\end{figure}


No additional devices can be installed near the four propellers of the Tello UAV, otherwise it will affect the power generated by the rotation of the UAV's propellers and threaten flight stability; There is an open source controller above the UAV's body, so there is no additional space for installing an extension bracket to fix the LED light source; light source can be installed in the connection frame between the UAV body and the propeller, but in order to ensure the balance of the UAV weight and avoid losing control during flight, it is necessary to install one light source at each of the connection frames connecting the four propellers. The total weight of the four LED light sources added together is about half of the UAV's own weight. Although the UAV can barely take off with this weight, the loss for the UAV greatly increases, the endurance and flight safety have been greatly affected, and due to the close proximity of the four light sources, the lighting range has not been significantly improved compared to a single light source. After comprehensive consideration, this installation approach is not recommended. Consequently, the location where an external light source can be installed is only below the fuselage. The rear part below the fuselage is the sensor of the UAV, responsible for the positioning and other functions of the UAV, so it cannot be blocked. The front part is the heat dissipation module of the UAV. If the light source is directly attached to the heat dissipation module, it will block most of the heat dissipation area, and the heat generated by the battery will not be well dissipated, resulting in danger. Therefore, the fixing method of the light source should be improved. A small insulating sponge is placed between the lamp source and the heat dissipation module of the UAV to avoid direct contact between them. After attaching the light source with two cable ties to the UAV, add two more cable ties to secure the cable ties holding the light source in place. In this way, the cable tie and light source will not easily shift, and the weight of the cable tie itself can be ignored. The most important is that this fixed mode has little effect on the heat dissipation module. The required components and installation renderings are shown in Fig. 4.


\begin{figure}[tbp]
\centering 
\includegraphics[scale=0.32]{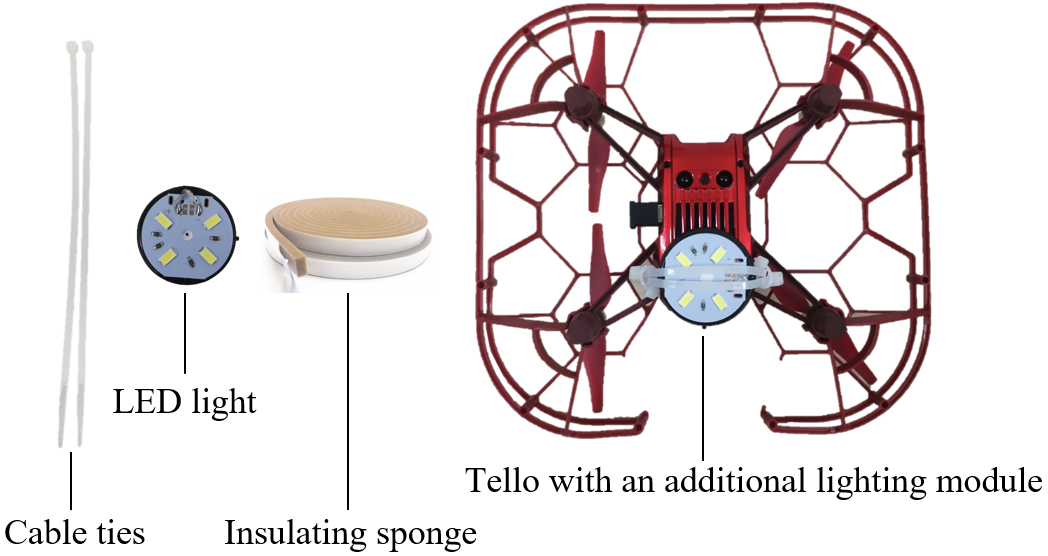}
\caption{Lighting Module Components and Installation Diagram}
\label{4}
\end{figure}

\subsection{Addressing Module Design}
\subsubsection{Model Building}
Based on the previously mentioned mathematical model, the modeling section creates a few functional models that must be used in the addressing module, including the model for calculating the distance between the UAV and the user, the model for calculating the UAV lighting range, the model for adjusting the UAV flight height, and the model for calculating the UAV power loss.

The calculation model for the distance between UAV and users can calculate the horizontal distance value between the current position of the UAV and all users. Two parameters are required: the UAV's position coordinates and the position coordinates of all users. The return value is the maximum distance between the UAV and all users.

The lighting range calculation model of a UAV can calculate the effective lighting range formed by the UAV at the current altitude. Two parameters, namely the flight altitude of the UAV and the effective lighting angle of the light source, need to be passed in, and the return value is the effective lighting range formed by the UAV at the current altitude.

The UAV flight altitude adjustment model can adjust the UAV's lighting altitude according to the actual situation after the UAV reaches the optimal deployment position. Three parameters need to be input: the UAV's flight altitude, the effective lighting angle of the light source, and the farthest distance between the user and the UAV. The return value is the adjusted flight altitude value of the UAV.

The power loss calculation model of UAV can calculate the service loss of UAV. The optimal deployment location coordinates of UAV and the user's lighting duration needs need to be input. The return values are flight power loss and total power loss.

\begin{figure}[tbp]
\centering 
\includegraphics[scale=0.32]{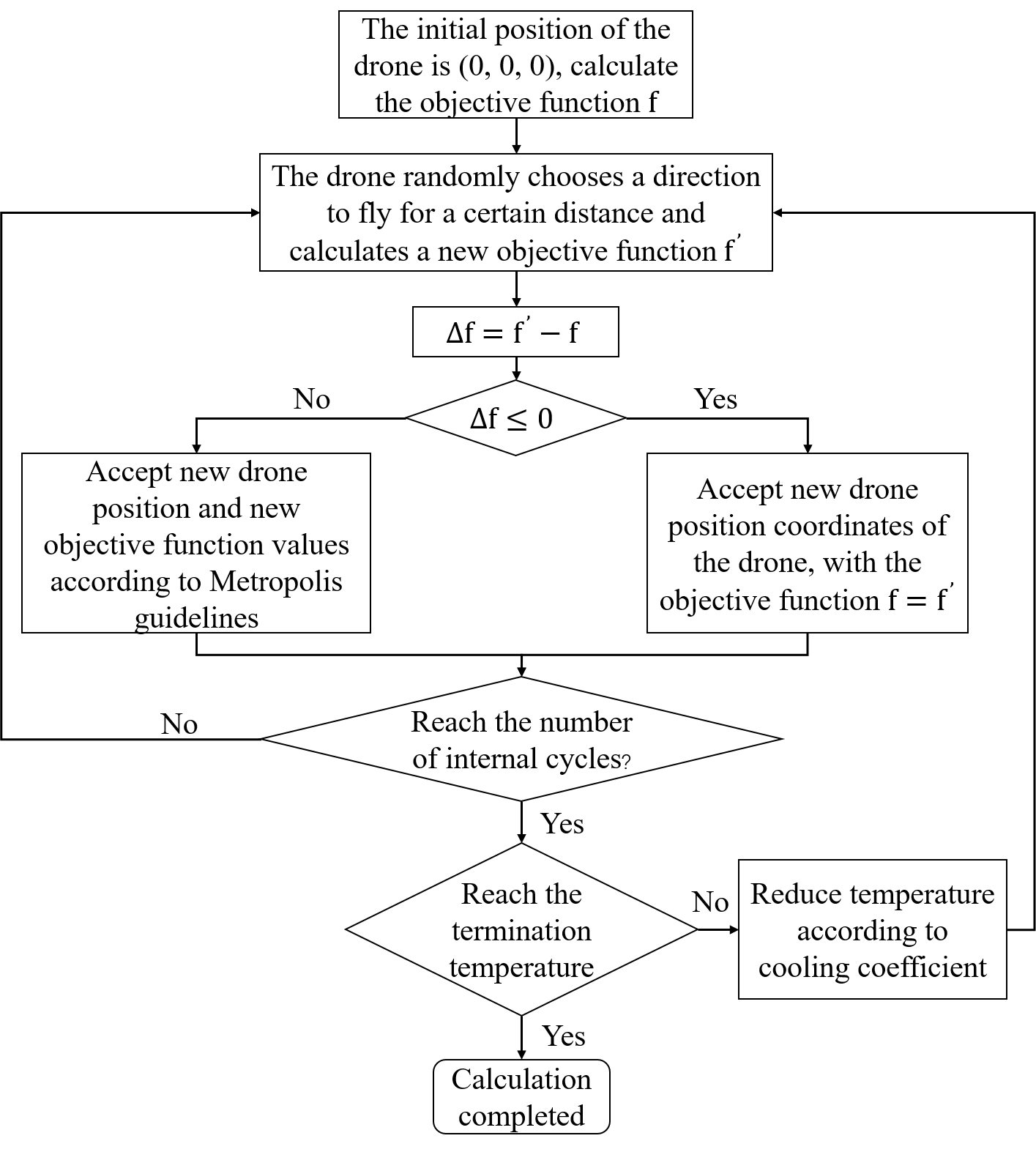}
\caption{Simulated Annealing Algorithm Iteration Flowchart}
\label{5}
\end{figure}

\subsubsection{UAV Addressing}
This paper divides user distribution scenarios into dense and sparse distribution scenarios based on the actual situation, and designs a simulated annealing algorithm to iteratively optimize the ideal deployment location of UAV. This is done because user distribution in real-world scenarios may be relatively concentrated, with a small overall range, or dispersed, resulting in larger continuous or discontinuous areas. Simulated Annealing is a global optimization algorithm based on probability. The idea of this algorithm comes from a process in solid state physics —— annealing process. Annealing is the process of heating a substance to a certain temperature and then slowly cooling it, causing atoms to rearrange to reach a lower energy state. Analogously, when solving a problem, the algorithm will search for the optimal solution in the initial high-energy state, and then slowly "cool down" to gradually reach the optimal solution. The basic idea of simulated annealing algorithm is to start from a random solution, accept a new or old solution in each iteration step, accept a worse solution than the current solution with a certain probability, and gradually reduce the probability of accepting a worse solution to avoid falling into a local optimal solution \cite{ait2022novel}. The algorithm controls randomness and the degree of greed in decision-making by controlling parameters. The specific design concept of simulated annealing addressing in this paper is shown in Fig. 5.


The optimization objective f of the algorithm is set as the maximum horizontal distance between all users and the UAV, with an initial temperature of 100, a termination temperature of 0.01, a cooling coefficient of 0.99, and 100 internal cycles. The acceptance criteria for Metropolis are set as shown in (10):
\begin{equation}
P = \left\{ \begin{array}{l}
1,\,\,\,\,\,\,\,\,\,\,\,\,\,\,\,\,\,\,\,\,\,\,\,\,\,\,\,\,\,\,\,\Delta f < 0\\\\
{e^{ - \frac{{f' - f}}{T}}},\,\,\,\,\,\,\,\,\,\,\,\,\,\,\,\,\Delta f > 0
\end{array} \right.
\end{equation}

\subsection{Flight Control Module Design}
\subsubsection{Control Command Framework}
Tello UAV can recognize specialized command and control statements, but there is no ready-made single command statement that can directly control UAV to complete a complex series of service actions. Therefore, in order to achieve the effect of automatic flight deployment of UAV, it is necessary to design a specialized command and control framework for UAV, select appropriate commands for combination and arrangement, and create flight text to record for subsequent reading and control by the system. Taking the cluster flight of two UAV as an example, the overall command framework designed in this article is shown in Fig. 6.
\begin{figure}[tbp]
\centering 
\includegraphics[scale=0.54]{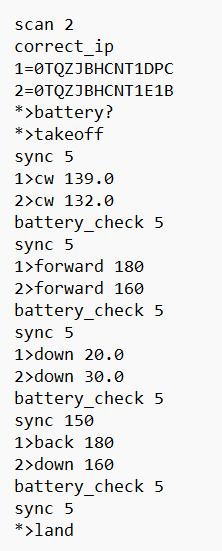}
\caption{Command Framework}
\label{6}
\end{figure}
The overall framework is divided into two parts. The first part realizes the search, connection and viewing of the relevant status of the UAV, while the second part controls the UAV to fly. In the figure, “scan2” to “$ *  > $takeoff” are the first part of the framework. The “scan2” command represents the need for the system to identify two UAV, that is, to search for and connect the corresponding number of UAV. The number here can be increased or decreased according to the actual situation. The “correct\_IP” command indicates that communication testing is required between the UAV and the system to ensure that both parties can receive information correctly. The “1=0TQZJBHCNT1DPC, 2=0TQZJBHCNT1E1B” command uses the SN code of each UAV to number and divide the UAV for subsequent command execution. The “$ *  > $battery?” command perform a battery check on all connected UAV, where “x$>$” indicates the execution of commands on “x” and “$ *  > $” indicates the execution of commands on all UAV. If any UAV's battery is in a dangerous state, it is not allowed to take off. After the above connection and check procedures are completed, execute the “$ *  > $takeoff” command to allow all UAV to take off. Afterwards, the UAV begins to execute the commands in the second part of the framework and enters the service cycle. The “cw” command controls the UAV's rotation angle to face the optimal deployment position. The “forward” command causes the UAV to move forward and deploy to the corresponding position. The “down” command controls the UAV to make height adjustments after reaching the deployment position. The "sync" command keeps the UAV in its current state, the back command causes the UAV to retreat and return, and the “land” command is a landing command, the “battery\_check 5” command is to check if the UAV's battery level will drop below 5\% during the process. The UAV will be compelled to land if its battery level drops to less than 5\% while flying in case it loses control.

\subsubsection{Create Flight Text}
Flight text is the key control text for achieving the automatic deployment effect of unmanned aerial vehicles. To ensure deployment effectiveness and flight safety, flight text needs to be developed in accordance with the command framework created earlier. After the addressing module calculates the optimal deployment coordinate point of the unmanned aerial vehicle, it is first necessary to calculate the parameter values required for automatic flight deployment of the unmanned aerial vehicle based on this coordinate, namely the required rotation angle, flight distance The height to be adjusted and the duration of providing lighting services, then create a flight text and write the relevant commands according to the command framework and fill in the parameters.

\begin{figure}[tbp]
\centering 
\includegraphics[scale=0.35]{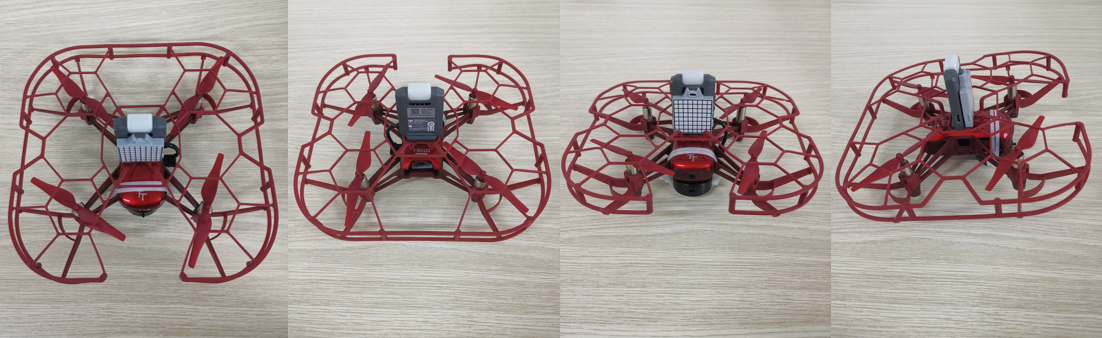}
\caption{Ground Placement Diagram after Adding Lighting Module}
\label{7}
\end{figure}

\subsubsection{UAV Flight Control}
The first step in UAV flight control requires the UAV to enter SDK mode and connect to the same server as the system, where SDK mode allows the Tello UAV to receive commands sent by the system and execute them. The specific operation first requires selecting the expansion module of the Tello UAV as the direct connection mode and turning on the power of the Tello UAV. Then, connect the UAV's WiFi on the PC terminal, and write a program using the “set\_AP” command transfers the name and password of the server that needs to be connected to the UAV and sends the “command: command” command to the UAV to enter SDK mode. If the UAV successfully provides feedback, it indicates that the UAV has entered SDK mode and can connect to the server. At this time, select the UAV's expansion module to network mode and restart, and the UAV can connect to the same server as the system, After the first basic work is completed, the system can start reading all the statements in the generated flight text and placing them in the system's command pool. Then, the command statements in the command pool are sequentially read and transmitted to the UAV. According to the designed hardware command framework, the system first searches for a corresponding number of UAV to connect and test communication between the two parties, and then numbers the UAV according to the SN code, After the numbering is completed, the UAV's battery level is checked. After the preparation work is completed, subsequent commands are executed. After each command statement is executed, the system waits for feedback from the UAV and determines whether the UAV has executed the corresponding command correctly. If it fails to respond correctly, the system will send the corresponding command to the UAV again so that the UAV can try to execute the command statement correctly again. If there is a flight control command for a single UAV during the command execution process, the system will not directly enter the waiting feedback state after reading and transmitting the command. Instead, it will continue to read all flight control commands for a single UAV in this section and send them to the corresponding UAV to ensure that all UAV can execute their own commands at the same time instead of queuing up for execution. Finally, the system will generate flight logs based on the status and command execution of the UAV during the flight process for subsequent review and analysis.


\begin{figure}[tbp]
\centering 
\includegraphics[scale=0.4]{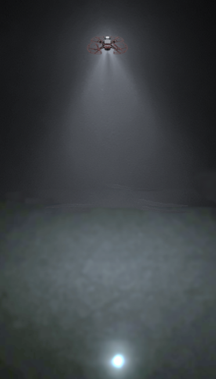}
\caption{Actual Lighting Effect of Lighting Module
}
\label{8}
\end{figure}

\section{System testing and results}\label{System testing and results}
\subsection{Lighting Module Test}
After installing the selected light source, the front half of the Tello UAV will be in a protruding state. However, after testing, the UAV can still be placed smoothly on a horizontal surface, as shown in Fig. 7. 



In actual flight tests, the UAV did not capsize during takeoff and landing, and hanging the LED light in the Tello's front half won't compromise its stability because of the light weight of the light. Similarly, an increase in load did not significantly slow down the UAV's flight speed. There is no discernible difference in the endurance performance of the Tello UAV after hanging LED lights, according to repeated comparison and testing.


After testing the Tello's added lighting module, it was discovered that the lighting effect is good and satisfies the design specifications. The shape of the area generated by lighting in the real scene is essentially consistent with the circular region in mathematical modeling. Fig. 8 displays the lighting module's measured impact.


Overall, the choice and installation technique of the LED lamp satisfy the established target requirements after actual flight and illumination testing, making it a viable approach for coupling the Tello UAV with an external lighting module.

\subsection{Addressing Module Test}
\subsubsection{Dense Scenes}
In this scenario, the distribution of users is rather concentrated, and the system bases its thorough analysis and decision-making on the distribution and demand data of users. To serve all users, only one UAV needs to be sent out. Based on the user information that was provided, the simulated annealing technique extracts the appropriate parameters for iteration; the final addressing iteration diagram is displayed in Fig. 9. The abscissa represents temperature and the ordinate represents the objective function f. As the temperature approaches the termination temperature, the objective function quickly decreases to the minimum value, and the algorithm converges stably. After practical testing, the UAV can provide services based on the calculated optimal deployment point, which can well meet the needs of all users. At the same time, after calculating the optimal deployment point of the UAV, the module calculates the power consumption of the whole service process combined with the lighting duration requirements of the user, and decides whether need a replacement UAV.

\begin{figure}[tbp]
\centering 
\includegraphics[scale=1.2]{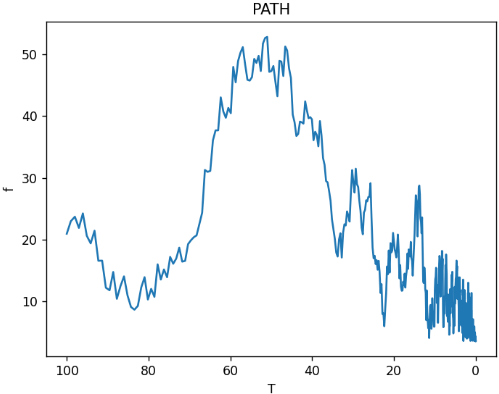}
\caption{Iterative Graph of Simulated Annealing in Dense Scenarios}
\label{9}
\end{figure}

\subsubsection{Sparse Scene}
In this scenario, the distribution of users is relatively scattered. After analyzing the distribution and needs of users, the system determines that a single UAV cannot meet the lighting task, and sends out a cluster of UAV for service. Whether it is a large-scale continuous area formed by the same demand user group or a discrete area formed by different demand user groups, the system can extract relevant parameters based on received user information and enter the simulated annealing algorithm to iteratively solve the optimal deployment location. Fig. 10 shows the simulated annealing addressing iteration diagram of two unmanned aerial vehicles serving at the same time, and it can be seen that the objective function f value tends to be optimal in the process of reaching the termination temperature, The iterative results of the deployment positions of the two UAV are stable and convergent. After practical testing, it has been found that in sparse scenarios, the UAV cluster can provide services based on the determined optimal deployment site, while still satisfying the needs of all users. At the same time, after determining the optimal position of each UAV in the UAV cluster to be deployed, the module combined with the lighting time required by the corresponding user to determine the power loss of each UAV in the process of providing service, and then determine whether replacement UAV is needed.

\begin{figure}[tbp]
\centering 
\includegraphics[scale=0.49]{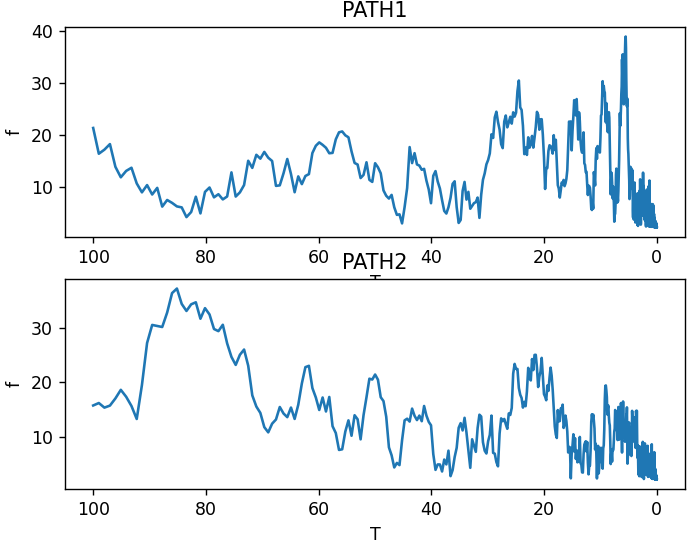}
\caption{Iterative Graph of Simulated Annealing in Sparse Scenarios}
\label{10}
\end{figure}

\subsection{Flight Control Module Test}
\subsubsection{Command Framework Testing}
The testing in this section was conducted under the previously designed command framework. Test the command framework by adding additional flight control command commands to the second part of the framework. The results indicate that under the designed command framework, the UAV can successfully establish a connection with the system and carry out a number of activities, including autonomous deployment and return under the control of complex and large-scale joint flight control commands. The overall command framework architecture can satisfy design expectations.

\subsubsection{Create Flight Text Test}
After iterating through the simulated annealing algorithm to find the optimal deployment location, this part can accurately calculate the rotation angle, flight distance, and height required for UAV deployment based on the position coordinate parameters. At the same time, it can create a new flight text at the specified location, write various commands, and fill in the calculated three parameters as well as the user's lighting duration requirements and other parameters. After inspection, it was discovered that the commands and parameters had been entered correctly, the overall flight text complied with the earlier-designed command framework, and the desired results had been obtained.

\begin{figure}[tbp]
\centering 
\includegraphics[scale=0.4]{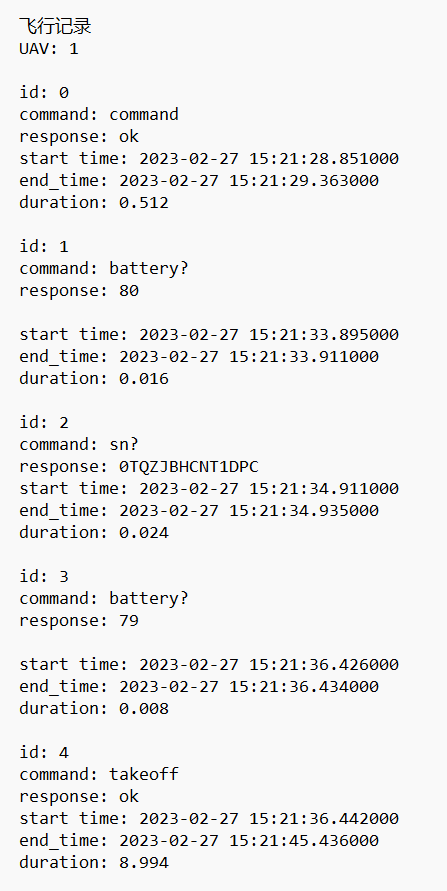}
\caption{Fragment of Flight Log}
\label{11}
\end{figure}

\begin{figure}[tbp]
\centering
\subfigure[]
{
        \includegraphics[scale=0.8]{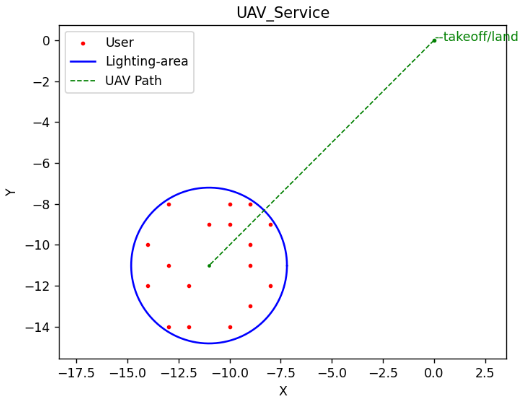}
}
\subfigure[]
{
        \centering
        \includegraphics[scale=0.9]{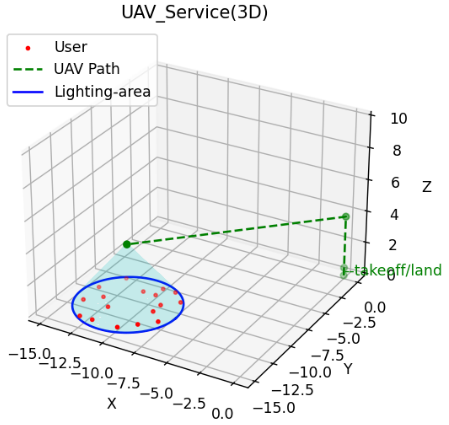}
}
\caption{Map of UAV Services in Dense Scenarios}
\end{figure}

\subsubsection{Flight Control Testing}
In multiple flight tests of the UAV flight control part, whether facing the control of a single UAV or a UAV cluster, the UAV can quickly establish communication with the system and maintain good communication status at all times. The transmission of commands and feedback from the UAV throughout the control process are timely and effective, and the number division and battery check commands of the UAV can be correctly executed to ensure subsequent safety. After reading the flight text and creating a command pool, the system can always execute according to the order of the commands in the flight text, without skipping commands or disorderly execution. In the case of a UAV cluster, the command execution synchronization of each UAV is also good. After the flight task is completed, the system can also automatically and correctly generate flight log text. The flight log style is partially depicted in Fig. 11.  Overall, this part can fly UAV automatically based on flight texts, and both the stability and deployment effect are good.


\subsection{Overall System Test Results}
The entire system was subjected to a thorough process test, and the test results showed that the designed emergency simulated lighting system can quickly respond to user requests in both dense and sparse scenarios, plan flight paths for UAV, and control their deployment services, effectively meeting the lighting needs of users.


The 2D schematic diagram of the UAV serving based on the calculated optimal deployment location in a dense scenario is shown in (a) of Fig. 12. The red dots represent the distributed users, and the UAV takes off from the labeled point and travels along the green dashed line to the calculated optimal target point for illumination. The blue circle represents the illuminated area of the UAV. After the service is completed, the UAV returns to the labeled point. (b) of Fig. 12 displays the more understandable 3D schematic diagram of the deployment service. After takeoff from the origin, the UAV will adjust its angle and raise it to a suitable height, fly to the user distribution area, and perform hovering lighting. The lighting effect is consistent with the mathematical modeling effect mentioned earlier, and can cover all users.


\begin{figure}[tbp]
\centering
\subfigure[]
{
        \includegraphics[scale=0.32]{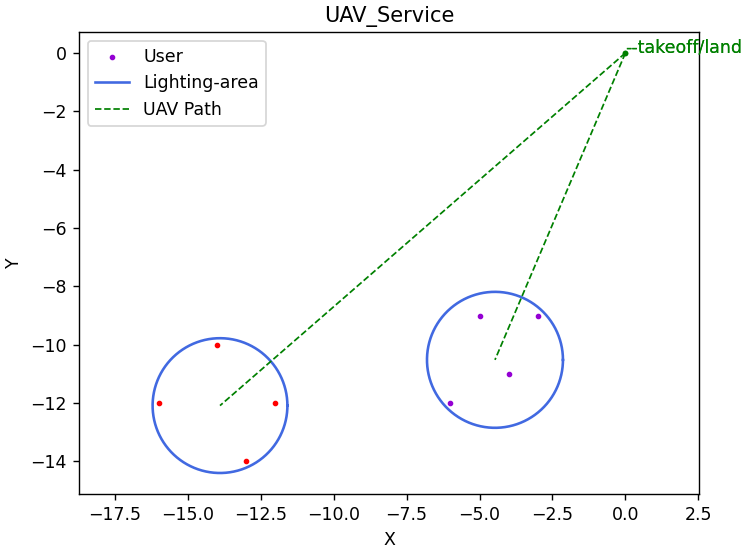}
}
\subfigure[]
{
        \includegraphics[scale=0.4]{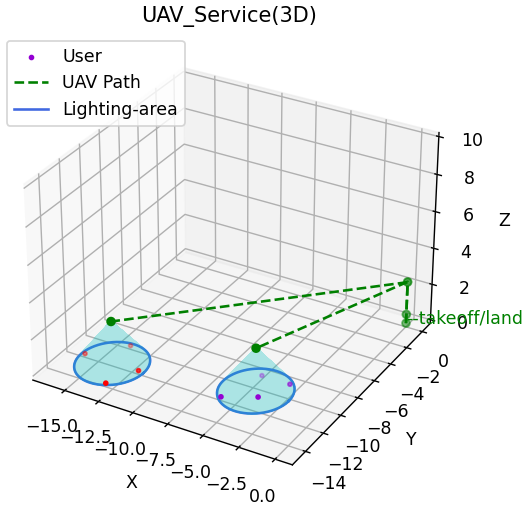}
}
\caption{Map of UAV Services for Different User Groups in Sparse Scenarios}
\end{figure}

(a) in Fig. 13 is a 2D schematic diagram of UAV services in a large-scale continuous area formed by the user of same demand in a sparse scenario. The red dots represent users, and it can be seen that all users in the figure are red, indicating that their lighting needs are consistent, but their distribution is relatively sparse, forming a continuous large-scale region. As shown in the figure, two UAV will fly along the green dashed line starting from the marked points, jointly form large-scale lighting to serve users in this area, and after the service is completed, the UAV cluster will return to the annotation point. (b) in Fig. 13 is a more intuitive three-dimensional schematic diagram. It can be seen from (b) that the large-scale lighting formed by the UAV cluster can cover all users at the same time.

\begin{figure}[tbp]
\centering
\subfigure[]
{
        \includegraphics[scale=0.35]{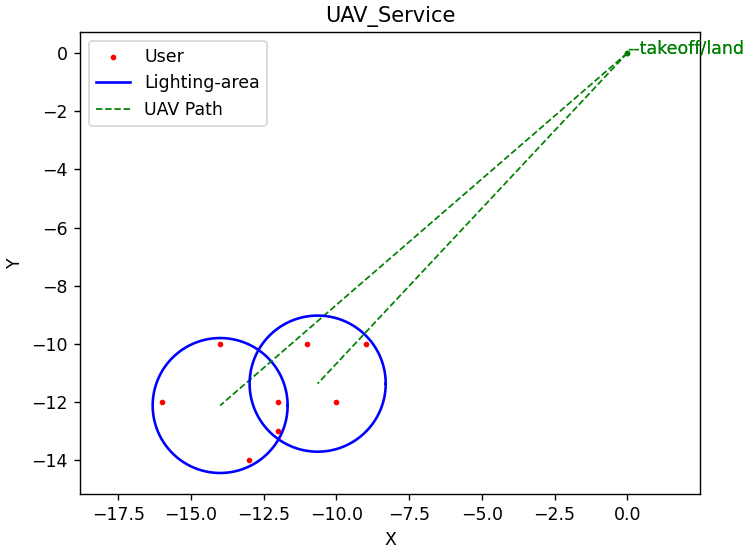}
}
\subfigure[]
{

        \includegraphics[scale=0.44]{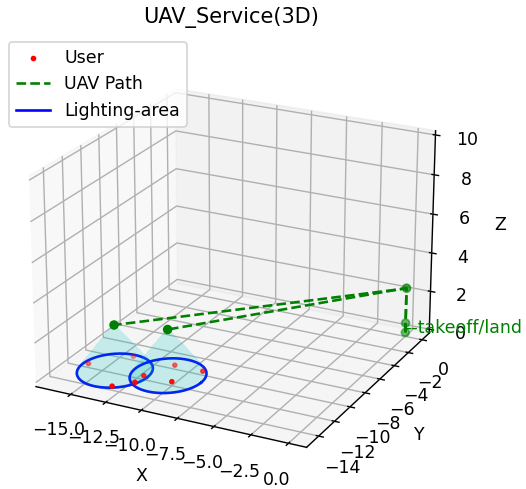}
}
\caption{Map of UAV Services for User Groups with Same Needs in Sparse Scenarios}
\end{figure}

\begin{figure}[tbp]
\centering 
\includegraphics[scale=0.48]{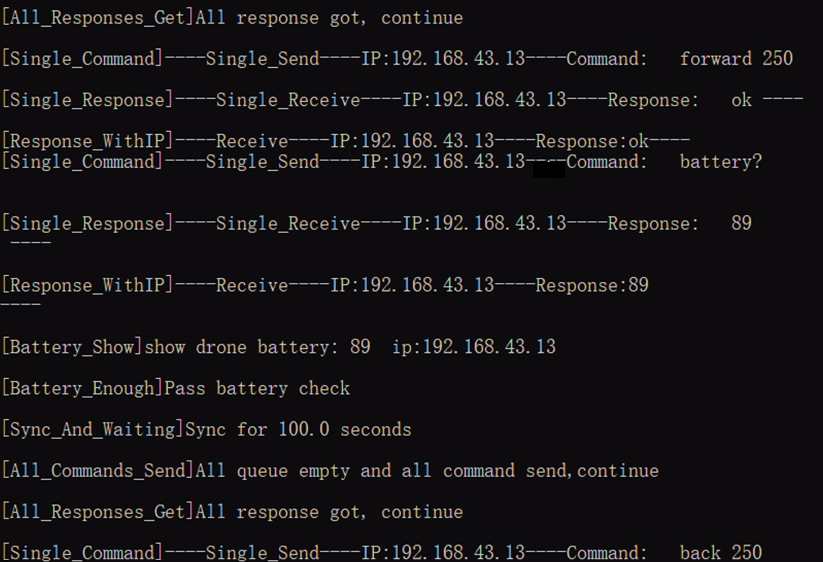}
\caption{Terminal Control and Feedback Interface}
\label{19}
\end{figure}

(a) in Fig. 14 shows a 2D schematic diagram of UAV services in a discrete area formed by different user groups with different needs in a sparse scenario. Red and purple dots represent user groups with different needs, which are distributed dispersedly and have inconsistent needs, forming two areas that are far apart. The UAV clusters will take off from the marked points and fly along the green dashed line to the corresponding user groups for illumination. After the service is completed, they will return to the marked points. (b) in Fig. 14 is a more intuitive three-dimensional schematic diagram.UAV can supply illumination services to various user groups based on their individual demands and can successfully cover all users.



Fig. 15 depicts the real evaluation of system performance in the event of a sandbox simulation of sudden power disruptions. Many users close to the Bird's Nest require emergency lighting assistance after an unexpected power outage. As a result, the emergency lighting system immediately assesses and determines user lighting requirements, scheduling UAV deployment accordingly. The total service effect is favorable.

\begin{figure}[tbp]
\centering 
\includegraphics[scale=0.365]{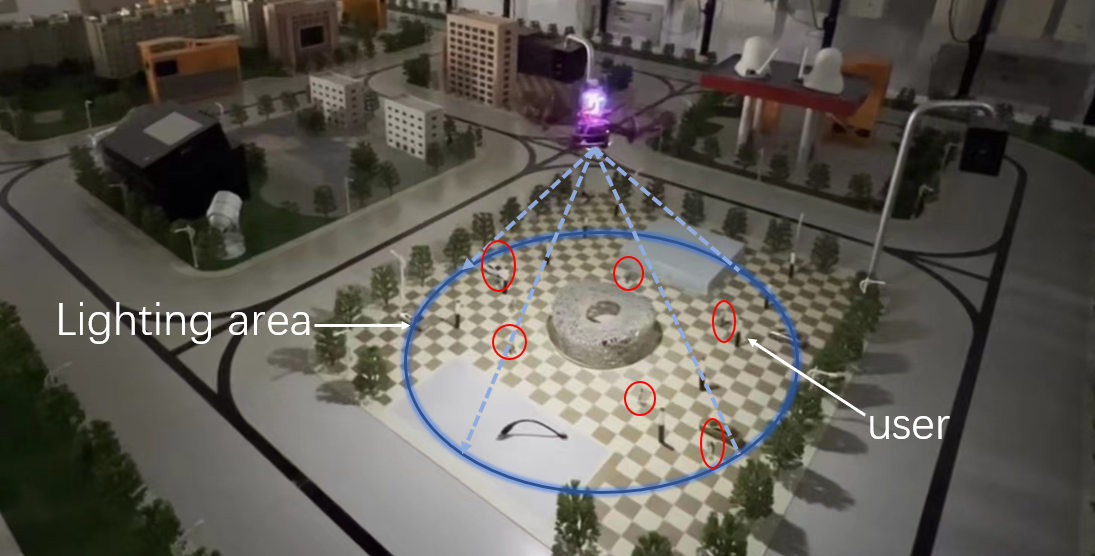}
\caption{Actual Measurement of UAV Service Effect}
\label{18}
\end{figure}

\begin{figure}[tbp]
\centering 
\includegraphics[scale=0.555]{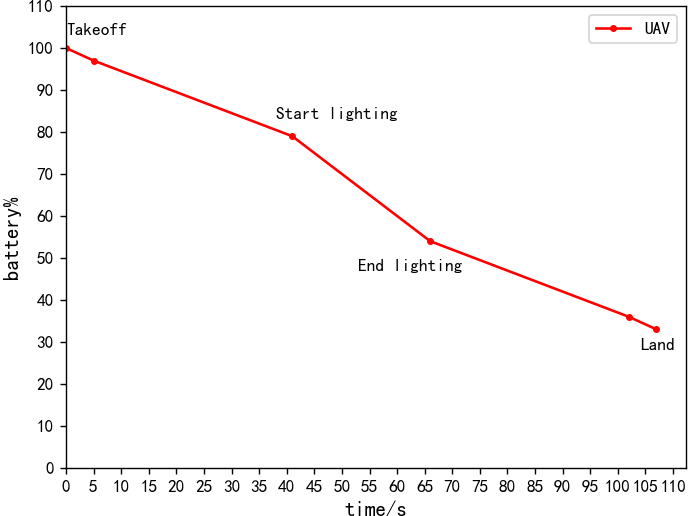}
\caption{UAV Battery Change Chart}
\label{20}
\end{figure}

Fig. 16 is a schematic diagram of the interaction interface between the system terminal and the UAV during flight deployment. When the system controls the UAV's flight, the system terminal can automatically send commands to the UAV, receive UAV responses, and provide status feedback on this interface.


After the system dispatches the UAV to complete a complete service cycle, the state feedback of the UAV at each time period will generate flight logs. Subsequently, the system extracts relevant state parameters during the UAV's flight process based on the flight logs and calculates them. Fig. 17 depicts a schematic illustration of the variations in electricity over the course of the full UAV service procedure.


\section{CONCLUSION}\label{CONCLUSION}
In this article, we present the development of an emergency simulation lighting system utilizing the Tello UAV. Our primary focus was on achieving quick and automated deployment of the UAV. The system design involved creating an external lighting source that aligns with the Tello UAV's characteristics. Additionally, we devised a simulated annealing algorithm to effectively plan the UAV's flight path. By integrating these components with a specially designed Tello command framework, the UAV can seamlessly collaborate with the system, executing various actions, including flight. The practical testing conducted confirmed the rationality and effectiveness of our comprehensive system design.

\section*{ACKNOWLEDGMENTS}
This paper was supported in part by the Zhejiang Provincial Natural Science Foundation of China (LZ23F010003, LQ23F010009), Zhejiang Provincial Key Laboratory of New Network Standards and Technologies (NNST)(No.2013E10012)，Zhejiang Gongshang University "Digital+" Disciplinary Construction Management Project (Project Number SZJ2022C010, SZJ2022A003).



\bibliographystyle{unsrt}
\bibliography{ref}
\end{CJK}
\end{document}